\documentclass[reprint,nofootinbib,amsmath,amssymb,fontenc aps]{revtex4-1}
\usepackage{graphicx,dcolumn,bm,hyperref,float}
\usepackage[mathlines]{lineno}

\newcommand{\beq}{\begin{equation}}
\newcommand{\eeq}{\end{equation}}
\newcommand{\bea}{\begin{eqnarray}}
\newcommand{\eea}{\end{eqnarray}}
\newcommand{\T}{\mathcal{T}}
\newcommand{\So}{\mathcal{S}}
\newcommand{\LGR}{\mathcal{L}_{\mbox{\tiny{GR}}}}
\newcommand{\Lkin}{\mathcal{L}_{\mbox{\tiny{kin}}}}
\newcommand{\Lint}{\mathcal{L}_{\mbox{\tiny{int}}}}
\newcommand{\Lex}{\mathcal{L}_{\mbox{\tiny{ex}}}}

\newcommand{\LGB}{\mathcal{L}_{\mbox{\tiny{GB}}}}
\newcommand{\Lag}{\mathcal{L}}
\newcommand{\ks}{\tilde{\kappa}}

\newcommand\supsetsim{\mathrel{\substack{\textstyle\supset\\[-0.2ex]\textstyle\sim}}}
\newcommand{\dt}{\frac{\partial}{\partial t}}

\begin{document}
	
\title{Symmetries from Locality. III. Massless Spin 2 Gravitons and Time Translations}

\author{Mark P.~Hertzberg}
\email{mark.hertzberg@tufts.edu}
\author{Jacob A.~Litterer}
\email{jacob.litterer@tufts.edu}
\affiliation{Institute of Cosmology, Department of Physics and Astronomy, Tufts University, Medford, MA 02155, USA
\looseness=-1}
	
\date{\today}
	
\begin{abstract}
We relax the assumption of time translation and Lorentz boost symmetry in theories involving massless spin 2 gravitons, while maintaining a basic notion of locality that there is no instantaneous signaling at a distance. We project out longitudinal modes, leaving only two degrees of freedom of the graviton. Our previous work, which assumed time translation symmetry, found that the Lorentz boost symmetry is required to ensure locality at leading order. In this work, without assuming time translations or Lorentz boosts, we show that locality of the exchange action between matter sources demands that massless spin 2, at leading order, organizes into Einstein-Hilbert plus a Gauss-Bonnet term with a prefactor that is constrained to be a particular function of time; while in the matter sector we recover time translation and Lorentz boost symmetry. Finally, we comment on whether the time dependence of the Gauss-Bonnet prefactor may be forbidden by going to higher order in the analysis and we mention that other possibilities are anticipated if graviton mass terms are included. 
\end{abstract}
	
\maketitle

\section{Introduction} 

The modern picture of cosmology is remarkably successful. However, it possesses a range of curious features; this includes the need for dark matter, dark energy, coincidences, fine-tuning problems, and, more recently, possible tension with different measurements of the Hubble parameter \cite{Verde:2019ivm}. It has sometimes been suggested that one or more of these issues may be addressed by modifying or deforming the underlying rules of gravitation, namely general relativity. 

But any theoretically sensible deformation of general relativity is difficult to come by. This is because there is a theorem \cite{Weinberg:1964ew,Weinberg:1965rz,Deser:1969wk}: {\em the unique local, unitary, Poincar\'e invariant theory of a species of massless spin 2 particles is, at large distances, general relativity.} (While multiple gravitons with higher order interactions can lead to inconsistencies \cite{Wald:1986bj,Hertzberg:2016djj,Hertzberg:2017abn}.) Hence to escape the conclusion that we necessarily have general relativity in a local and unitary way, one must do one (or more) of the following: (i) consider small distances, (ii) add extra degrees of freedom, (iii) deform Poincar\'e symmetry. The first option (i) is not interesting to resolve puzzles in cosmology, which involve phenomena on the largest of scales; so we shall not pursue this possibility here. In the second option (ii) one may (a) add a mass to the graviton (which takes the number of degrees of freedom from 2 helicities to 5 polarizations). However, this possesses various strong coupling problems and no known UV completion (arguably due to problems with causality), so we shall also ignore this possibility here. Or one may (b) add other types of particles, especially spin 0 particles, which can potentially mediate long range forces if they are extremely light. Such interactions are tightly constrained by solar system tests and are unlikely to solve the above conceptual problems anyhow (one can just add scalars as part of the matter sector to act as dark matter of course); so we shall not pursue this possibility here either. 

This leaves us with investigating option (iii), which has been investigated in the literature \cite{AmelinoCamelia:2000mn,Magueijo:2001cr,Kostelecky:2003fs,Collins:2004bp,Mattingly:2005re,Horava:2009uw,Khoury:2013oqa,Collins:2006bw,Charmousis:2009tc,Papazoglou:2009fj,Griffin:2017wvh,Moffat:1992ud,Pajer:2020wnj}, and is the focus of this paper. In our previous papers \cite{Hertzberg:2017nzl,part1,part2} we considered deforming Lorentz boosts, while maintaining all other parts of the Poincar\'e symmetry, namely rotation symmetry, spatial translation symmetry, and time translation symmetry. In \cite{part1} we considered spin 1 particles (electromagnetism) and in \cite{part2} we considered spin 2 particles (gravitation). By imposing a very basic notion of locality, that there is no instantaneous signaling, we showed that these particles must couple to conserved sources, associated with a symmetry. In the case of spin 1 we showed it must couple to a conserved current, which requires the familiar $U(1)$ symmetry, and for spin 2 we showed it must couple to a conserved 2-index symmetric current (energy-momentum tensor), which requires Lorentz boost symmetry. Since the latter recovers Lorentz boosts and in turn the full Poincar\'e symmetry involving spin 2, then it is anticipated to recover the full structure of general relativity just from locality of spin 2.

For potential cosmological applications it is important to move beyond the above assumptions. Most importantly, cosmology involves an expanding universe, i.e., a background space-time that breaks time translation invariance. While in the context of general relativity this is associated with spontaneous breakdown of time translations, there remains several mysteries as mentioned earlier. Therefore it motivates exploring setups in which time translation symmetry is explicitly broken. This work is a first step in this direction. Our particular set-up is as follows: we will assume, again, a preferred frame with rotation invariance (which allows for a notion of particle spin) and translation invariance in space, but we will not assume Lorentz boosts or time translation invariance. As a first investigation into this problem, we will assume the graviton is massless. This may seem a very mild assumption since we are going to project down to only the 2 polarizations of the spin 2 graviton and we know that Poincar\'e symmetry plus unitarity would dictate that it would need to be massless. But since we are not assuming Poincar\'e symmetry in this work, this argument is no longer valid. In fact in our previous paper \cite{part2} where we did not assume Lorentz boosts, we allowed for mass terms for the graviton, but went on to show that locality required them to vanish. If we now give up both Lorentz boost and time translation invariance, we can again have mass terms for the graviton, and for the greatest generality we should only use some principle like locality to dictate what terms are allowed. However, it turns out this issue is rather complicated. So, for simplicity, we will assume the graviton is massless here, and leave the inclusion of mass terms for future work. We note that all current data is indeed consistent with our assumption of a massless graviton \cite{Tanabashi:2018oca}. However, we know that when expanding around a Friedmann-Robertson-Walker (FRW) background space-time in standard general relativity, mass terms appear; so our work here will not be able to recover perturbations around FRW. Instead we view the present work as a solid starting point to explore the explicit breakdown of time translations, albeit not the most general. And to be clear, we are especially interested in explicit breakdown, rather than spontaneous, which is already a well studied problem.  

The structure of this paper is similar to the previous papers \cite{part1,part2}, especially to the gravitation paper \cite{part2}, and we encourage readers to read these accompanying papers. Since many details are already laid out there, our description here will be relatively brief, with a focus on the construction and primary results. 
Our paper is organized as follows: We firstly recap the fact that standard general relativity is local. We then discuss a family of generalizations that break time translations and Lorentz boosts. We then focus on classes of theories that cut down to 2 polarizations of the graviton. In each case we impose locality, defined here as avoiding instantaneous signaling, and find the same resulting theory. We believe this resulting theory is valid regardless of the particular choice of how to cut down the degrees of freedom. We find that we recover general relativity (GR) to the leading order we are working, plus an additional set of terms that involve a very specific form of time translation symmetry violation. We then comment on what is anticipated to occur when working to higher order.

\section{Theories of Massless Spin 2}\label{spin2theories}

\subsection*{Recap of Poincar\'e Invariant Massless Spin 2}

Our goal in this work is to study and generalize theories of massless spin 2 particles. We will begin with the standard theory of general relativity, defined around a flat background, and then generalize this in a particular fashion that we will describe in the next subsection. So it is useful to recap the structure of general relativity (the unique local, unitary, Poincar\'e invariant theory of massless spin 2). To study the leading order interactions, we expand the metric $g_{\mu\nu}$ around a flat background as
\beq
g_{\mu\nu} = \eta_{\mu\nu}+\kappa\,h_{\mu\nu}
\eeq
where $\eta_{\mu\nu}$ is the metric of Minkowski space-time and $h_{\mu\nu}$ is the spin 2 field whose purpose is to describe interactions of spin 2 particles in a local way. To  project out the unphysical degrees of freedom, one needs the gauge redundancy: $h_{\mu\nu}\to h_{\mu\nu}+\partial_{(\mu}\alpha_{\nu)}$. It is well known that this leads to the unique Poincar\'e invariant quadratic kinetic term for gravitons, which is the Einstein-Hilbert term expanded to quadratic order (up to field redefinitions and boundary terms) 
\bea
\Lkin&=&\frac{1}{2} \left( \eta^{\alpha\beta} \partial_\alpha h^{\mu\nu} \partial_\beta h_{\mu\nu} - \eta^{\mu\nu} \partial_\mu h^{(4)} \partial_\nu h^{(4)} \right)\nonumber\\
&+&\partial_\mu h^{\mu\nu} \partial_\nu h^{(4)} - \partial_\mu h^{\mu\alpha} \partial_\nu h^\nu_\alpha 
\label{LGR}\eea
where $h^{(4)}$ is the 4-dimensional trace $h^{(4)}=\eta^{\mu\nu} h_{\mu\nu}$. Furthermore, in order to maintain the gauge invariance at leading order, the graviton must couple to a symmetric 2-index conserved current, the energy-momentum tensor, as $\Lint =-{1\over2} \kappa\, h_{\mu\nu} T^{\mu\nu}$. Here the coupling $\kappa$ is $\kappa\equiv\sqrt{32\pi G_N}$, where $G_N$ is Newton's constant. To quadratic order in $h_{\mu\nu}$, the full action is $\LGR^{(2)}=\Lkin+\Lint$. 

To leading order, i.e., ignoring the induced influence of gravitation on the matter sector itself, $T^{\mu\nu}$ obeys $\partial_\mu T^{\mu\nu}=0$. Since the graviton is massless and mediates long ranged forces, it is highly non-trivial that it leads to local interactions (we note that just because a theory is Lorentz invariant does {\em not} imply locality; it is trivial to construct Lorentz invariant theories that are non-local. For example, consider the Lagrangian for a scalar field $\mathcal{L}=(\partial\phi)^2/2-(\partial\phi)^4/\Lambda^4+\ldots$; this leads to superluminality around non-trivial $\phi$ backgrounds \cite{Adams:2006sv}). In fact it can be shown that if the sources were to disobey the null energy condition $T_{\mu\nu}n^\mu n^\nu\geq0$, then in fact the graviton could itself be superluminal around certain kinds of background configurations. Instead if we assume the null energy condition, the theory of general relativity, with standard leading order interactions, is in fact local. 

As an important manifestation of this, we can consider the tree-level graviton exchange between a pair of matter sources. In harmonic (de Donder) gauge $\partial_\mu h^\mu_\nu={1\over2}\partial_\nu h^{(4)}$ the equations of motion are known to become very simple
$\square h_{\mu\nu}=-{\kappa\over2}\left(T_{\mu\nu}-{1\over2}T^{(4)} \eta_{\mu\nu}\right)$, 
with $\square=\partial_t^2-\nabla^2$. 
One can then readily check if the tree-level exchange action between sources is local. To do so, one first solves for the particular solutions $h_{\mu\nu}=-{\kappa\over2\square}\left(T_{\mu\nu}-{1\over2}T^{(4)} \eta_{\mu\nu}\right)$. The corresponding tree-level graviton exchange action between sources is half the interaction term $-{1\over4}\kappa\,h_{\mu\nu} T^{\mu\nu}$, giving the result
\beq
{8\Lex\over\kappa^2} =T_{ij} \frac{T_{ij}}{\square} - \frac{T}{2} \frac{T}{\square} + \frac{T_{00}}{2} \frac{T_{00}}{\square} - 2 T_{0i} \frac{T_{0i}}{\square} + T_{00} \frac{T}{\square} \,\,\,\,\,\,
\label{Lint_GR}\eeq
Here we have broken up the energy-momentum tensor into a spatial tensor $T_{ij}$, spatial vector $T_{0i}=T_{i0}$, and spatial scalar $T_{00}$ (along with $T\equiv\delta_{ij}T^{ij}$). This decomposition will be useful for comparison to our later analyses where we break Lorentz symmetry explicitly. This exchange action is local, despite the presence of  the inverse operators on the right hand side. That is because the inverse operators are the wave-operator, which is associated with retarded wave propagation, as opposed to the inverse Laplacian, which is associated with instantaneous action at a distance. Later when we deform several aspects of the Poincar\'e symmetry, we shall see that almost all deformations will in fact lead to inverse Laplacians and instantaneous action at a distance. So by imposing locality, we will be able to derive that almost all deformations are forbidden, as we saw in our previous pair of papers \cite{part1,part2}.

\subsection*{Spatially Invariant Massless Spin 2}

Let us make clear what symmetries we will explicitly break and which remain. We assume a preferred frame that has rotation invariance and spatial translation invariance. However, as in our previous pair of papers \cite{part1,part2}, we do not assume Lorentz boosts. Moreover, we now do not assume time translations. Maintaining rotation will be important in this work. In particular, this ensures we have a notion of particle spin and our goal will be to build an interacting theory of spin 2 particles. Since we are interested in maintaining an explicit notion of locality, we will use the formalism that makes locality manifest; namely we can embed the spin 2 particles into fields. In principle we only need a $3\times3$ symmetric matrix to describe this, which we call $h_{ij}$, which transforms as a tensor under rotations. But locality will in fact require us to also introduce a non-dynamical scalar $\phi$ and a non-dynamical vector $\psi_i$. In the Lorentz invariant case, these can be organized into a $4\times4$ symmetric matrix field $h_{\mu\nu}$ that transforms as a tensor under Lorentz transformations (up to gauge transformations). In the non-Lorentz invariant, but rotationally invariant, case we can just refer to its components as
\beq
h_{00}\equiv \phi,\,\,\,\,\,\,\,\,h_{0i}=h_{i0}\equiv \psi_i,\,\,\,\,\,\,\,\, h_{ij}
\eeq
(where $\phi$ and the Newtonian potential $\phi_N$ are related by $\phi=2\phi_N/\kappa$). Also, the 3-trace is $h\equiv\delta^{ij} h_{ij}$. 

To leading order, we need to form some appropriate generalization of the above dimension 4 operators that describe the quadratic terms of eq.~(\ref{LGR}). In this work we are giving up both Lorentz boosts and time translations; this means there are many ways we can generalize the above theory of Poincar\'e invariant massless spin 2. A concrete generalization is the following: we can write out each of the terms of eq.~(\ref{LGR}) in terms of its rotationally covariant building blocks $h_{ij}$, $\psi_i=h_{0i}$, and $\phi=h_{00}$. Then we can insert a different function in front of every term as follows
\bea
&\Lag&= -2A\,\dot{\psi}_i \partial_j h_{ij} + 2B\, \dot{h} \partial_i \psi_i  -C\, \partial_i \phi \partial_i h + D\, \partial_i \phi \partial_j h_{ij} \nonumber\\
&-& E\, \partial_i h \partial_j h_{ij} - F \left( \partial_i \psi_i \right)^2 + G\, \partial_j h_{ij} \partial_k h_{ik}  + H\,\partial_j \psi_i \partial_j \psi_i  \nonumber\\ &-& {I\over2} \dot{h}^2 + {J\over2} \partial_i h \partial_i h + {K\over2} \dot{h}_{ij} \dot{h}_{ij} - {L\over2} \partial_k h_{ij} \partial_k h_{ij}  + \Lint
\label{General}\eea
where the coefficients are allowed to {\em a priori} be arbitrary functions of time, $A(t),\,B(t),\ldots,\,L(t)$. The Poincar\'e invariant theory of massless spin 2 is recovered by setting all these coefficients to 1. Although the above 12 kinetic terms are the full set of dimension 4 terms that describe the graviton in the Poincar\'e invariant case, it is not the full set allowed if we give up Lorentz and time translation symmetry as we are doing here. In particular one can also add a pair of terms $\sim(\partial_i\phi)^2$ and $\sim\dot\phi\,\partial_i\psi_i$. However it is relatively simple to show (as we mention in \cite{part2}) that they lead to non-locality, so they shall be ignored here. 

There are lower dimension quadratic terms that could be included too. In particular one could include several different dimension 2 mass terms and dimension 3 single-derivative terms. In the time translationally invariant theory that we analyzed in Ref.~\cite{part2}, we addressed these possible terms. In particular we allowed for mass terms, at least in part, and we mentioned that one could then perform field re-definitions to remove the dimension 3 terms (generating dimension 5 terms that are higher order and ignorable in this analysis); we then showed that mass terms led to non-locality. Similarly, here we could include such lower dimension terms. Indeed this is important to obtain perturbations around FRW in general relativity. However, for simplicity, this will not be our focus here. Instead we will focus on massless spin 2 and assume the leading quadratic terms are dimension 4, as they are in general relativity. We suspect that if one sets the masses to zero then in fact the dimension 3 terms must vanish in order to maintain locality, though we do not have a proof to present here. We leave the important case of including time dependent mass terms and time dependent dimension 3 terms to future work. 

The leading order interaction allowed involves a single graviton coupled to matter. Poincar\'e invariance requires the coupling to the energy-momentum tensor $\sim h_{\mu\nu} T^{\mu\nu}$, as we reviewed in the previous subsection. We generalize this to
\beq
\Lint =-{1\over2} \kappa \, h_{\mu\nu} \T^{\mu\nu} \label{Lint}
\eeq
where $\T^{\mu\nu}$ is a general source which does not {\em a priori} have any relationship to an energy-momentum tensor. In fact since we do not {\em a priori} assume Lorentz boosts or time translation invariance there does not {\em a priori} even exist a conserved, not to mention symmetric, energy-momentum tensor. We note that we will again specify the coupling as $\kappa$, which is useful for power counting. One could promote the coupling to be a function of time $\kappa=\kappa(t)$; however, without loss of generality, we can absorb this time dependence into the sources instead. Since we are only assuming spatial rotation invariance, we again find it most convenient to decompose it into its scalar, vector, and tensor (under rotations) pieces, as
\beq
\T^{00}\equiv \rho,  ~~~~ \T^{0i} = \T^{i0} \equiv p_i, ~~~~ \T^{ij} \equiv \tau_{ij}
\label{Tcomponents}\eeq

This class of theories of massless spin 2 is specified by 12 functions of time $A(t),B(t),\ldots,L(t)$ (in addition to the strength of gravity $\kappa$). As mentioned above, the Lorentz invariant and time translationally invariant limit (general relativity) is where all these quantities are set to 1. Hence we have a large space of deformations from general relativity; namely 12 new functions of time. However, several of these functions can be eliminated by re-defining the fields $\phi$, $\psi_i$, and $h_{ij}$ and exploiting the degeneracy in meaning between the scalar sources $\rho$ and the trace of $\tau_{ij}$; this leaves 8 independent functions.

The classical equations of motion that follow from the action (\ref{General}) are 
\bea
\ks \, \rho &=& -C \nabla^2 h + D \partial_i \partial_j h_{ij} \label{equations of motion1} \\
\ks \, p_i &=&  B \partial_i \dot{h} + H \nabla^2 \psi_i -F \partial_i \partial_j \psi_j- {\partial \over \partial t} \left( A \, \partial_j h_{ij} \right) \label{equations of motion2} \\
\ks \, \tau_{ij} &=&  2 \delta_{ij} {\partial \over \partial t} \left( B \, \partial_k \psi_k \right) - \delta_{ij}  {\partial \over \partial t} \left( I \dot{h} \right) +  {\partial \over \partial t} \left( K \dot{h}_{ij} \right) \nonumber\\ &-&A \partial_{(i} \dot{\psi}_{j)}  
+ D \partial_i \partial_j \phi - E \partial_i \partial_j h - E \delta_{ij} \partial_k \partial_l h_{kl} \nonumber\\ &-& C \delta_{ij} \nabla^2 \phi 
+ G \partial_k \partial_{(i}h_{j)k}+J \delta_{ij} \nabla^2 h - L \nabla^2 h_{ij}\,\,\,\,\,\,\,\,\,\,\,\,
\label{equations of motion}
\eea
where $\ks\equiv-\kappa/2$. We can see breaking time translation symmetry results in one additional term in eq.~(\ref{equations of motion2}) proportional to $\dot{A}$ and three additional terms in eq.~(\ref{equations of motion}) proportional to $\dot{B},\,\dot{I},\,\dot{K}$, as compared to GR in which $A=B=\mathellipsis=L=1$ are all constants.

The sources $\T^{\mu\nu}$ are not obviously required to be conserved here. As we did in \cite{part2}, we parameterize the violation of source conservation by a scalar $\sigma$ and vector $w_i$ as
\bea
\sigma &\equiv& \partial_i p_i + \dot{\rho}_r \\
w_i &\equiv& \partial_j \tau_{ij} +  \dot{p}_{r,i}
\eea
where $\dot{\rho}_r \equiv (A/D) \dot{\rho}$ and $\dot{p}_{r,i} \equiv (A/H) \dot{p}_{r,i}$. In the GR limit we have $\partial_\mu\T^{\mu\nu}=0$, which means $\sigma=w_i=0$, but these are generally non-zero when Lorentz and time translation symmetry is abandoned. In fact we can use the classical equations of motion to express these violations of conservation in terms of the fields as
\bea
\ks\, \sigma &=&  (H-F) \nabla^2 \partial_i \psi_i + \left( B - \frac{A C}{D} \right) \nabla^2 \dot{h} \nonumber \\
 &+& \left( A {\dot{D} \over D} - \dot{A} \right) \partial_i \partial_j h_{ij} - {A \over D} \dot{C} \nabla^2 h     \label{sigmaEq}\\
\ks\, w_i &=& \left( 2B - A - \frac{AF}{H} \right) \partial_i \partial_j \dot{\psi}_j + 2\dot{B} \partial_i \partial_j \psi_j \nonumber \\
&+& \left( \frac{AB}{H} - I \right) \partial_i \ddot{h} +\left( {A \dot{B} \over H} - \dot{I} \right) \partial_i \dot{h} \nonumber\\
&+& \left( J - E \right) \partial_i \nabla^2 h +  \left( G - L \right) \nabla^2 \partial_j h_{ij} - {A \ddot{A} \over H} \partial_j h_{ij} \nonumber\\
&+& \left( K - \frac{A^2}{H} \right) \partial_j \ddot{h}_{ij} + \left( \dot{K} - 2 {A \dot{A} \over H} \right) \partial_j \dot{h}_{ij}   \nonumber \\
&+& \left( G - E \right) \partial_i \partial_j \partial_k h_{jk} + \left(D - C \right) \partial_i \nabla^2 \phi \,\,\label{wEq}
\eea
This representation is useful because it shows explicitly that $\sigma$ and $w_i$ vanish in the GR limit $A=B=\ldots=L=1$, since all terms on the right hand side are zero, but are non-zero otherwise.

\section{Restricting to Two Degrees of Freedom}

As is well known, in the Poincar\'e invariant case the massless unitary representations are individual helicities, and for spin 2 we need to keep both left and right handed helicities to build a theory with local interactions. Furthermore, this uniquely specifies the free theory of eq.~(\ref{LGR}) as it carries the gauge redundancy needed to cut down to these 2 degrees of freedom: $h_{\mu\nu}\to h_{\mu\nu}+\partial_{(\mu}\alpha_{\nu)}$. Then one can construct all the interactions, order by order in $\kappa$, to build up the full Einstein-Hilbert action. 

In contrast, it is sometimes suggested that the existence of the graviton can be {\em derived} by promoting the global Poincar\'e symmetry to the gauge symmetry of diffeomorphisms. But this is incorrect, as any gauge symmetry can be trivially introduced by the Stueckelberg trick. Instead, the only known way to introduce the graviton within the framework of effective field theory is to simply {\em postulate} it.

Now in the present work, since we are giving up Lorentz boost and time translation invariance, but still studying massless spin 2, there is no longer an unambiguous consequence for the number of degrees of freedom. However, we will once again just postulate the existence of the 2 polarizations of the graviton in this work. We are motivated by the following: (i) we wish to make only small deformations to general relativity, while additional degrees of freedom could be viewed as large changes; (ii) all current evidence of gravitational waves are consistent with only 2 degrees of freedom \cite{Abbott:2016blz}; and (iii) in the Lorentz invariant case additional degrees of freedom in massive gravity have been argued to suffer strong coupling and causality problems \cite{ArkaniHamed:2002sp,Adams:2006sv,Deser:2012qx,Deser:2014hga}. 

Since the theory of eq.~(\ref{General}) with pre-factor functions $A(t),\,B(t),\,\ldots,\,L(t)$ that are {\em a priori} allowed to be arbitrary functions of time, there is no longer any gauge redundancy in the theory. This means that it can propagate more than 2 degrees of freedom. Hence to project down to only 2 degrees of freedom for the graviton, one has to impose some constraints on the fields and/or the pre-factor functions. 
%In our previous work \cite{part2} we described three ways to do this, which were essentially the most general set of possibilities. Furthermore, the 3rd option was readily shown to be a special case of the 2nd option once some basic notion of locality was enforced. Hence we really need only focus on the first two options, as we shall do here: 
We consider two options, which encompass the most general set of possibilities: (A) Set $\partial_i \psi_i = 0 = \partial_j h_{ij}$; (B) Fix parameters such that $\partial_i \psi_i$ and  $\partial_j h_{ij}$ are directly determined by the equations of motion. Both of these examples may seem like a form of ``gauge fixing", but we emphasize that since there is no gauge redundancy assumed, they are instead a choice of theory with different physical predictions. In our previous work of Ref.$~$\cite{part2} we considered a third option as well. However, after enforcing locality, this third option was readily seen to be a special case of the second, so we will not consider it here.

Nevertheless, by demanding the tree-level exchange action is local, we find at the leading order to which we are working that both cases (A) and (B) will collapse to the same theory; namely GR with some additional terms. As in our previous work, we also do not {\em a priori} impose conservation laws on the sources $\T^{\mu\nu}$. However, locality will restrict its form tremendously, as we will describe.

\subsection*{Theory A: Transverse Constraint}\label{grcase1}	

In this and the next section we describe two explicit ways to cut down to 2 degrees of freedom in the above class of theories. This is analogous to the first 2 theories we studied in \cite{part2} (while the third theory studied in \cite{part2} was found to be only a special case of the second theory once locality was imposed, so we won't repeat that analysis here). Moreover, we believe that our results are general and apply to any such choice.

Perhaps the most basic way is to remove any longitudinal modes directly by imposing that they vanish as
\beq
\partial_i \psi_i = 0,\,\,\,\,\,\,\,\,\,\,\partial_j h_{ij} = 0  \label{gauge}
\eeq
Note that this choice is not merely a ``gauge choice", since the lack of Lorentz symmetry in the theory means there is no gauge redundancy in our field formalism, but a choice of theory. We will only need to work with the classical theory in this analysis (which suffices to derive our basic conclusions) and so we can imagine the above is implemented by Lagrange multipliers. The corresponding classical equations of motion are
\bea
\ks \, \rho &=& -C \nabla^2 h  \label{rhoA}\\
\ks \, p_i &=& B \partial_i \dot{h} + H \nabla^2 \psi_i \label{pA} \\
\ks \, \tau_{ij} &=&  - \delta_{ij} {\partial \over \partial t} \left( I \dot{h} \right) + {\partial \over \partial t} \left( K \dot{h}_{ij} \right) - L \nabla^2 h_{ij} -A \partial_{(i} \dot{\psi}_{j)}\nonumber\\
 &-& C \delta_{ij} \nabla^2 \phi + D \partial_i \partial_j \phi - E \partial_i \partial_j h  + J \delta_{ij} \nabla^2 h  \,\,\,\,\label{tauA}
\eea
Note the functions $F$ and $G$ no longer appear in the theory. 
	
We can explicitly check on the status of source conservation in this theory. The scalar component of the usual conservation of sources $\partial_\mu T^{\mu\nu}=0$ is now
\beq
\partial_i p_i + B\,  {\partial \over \partial t} \!\left( {\rho \over C} \right) = 0 \label{cons0}
\eeq
so the quantity $\int d^3\rho/C$ is conserved, though in our choice of variables it is not simply the integral of the source $\rho$, but re-scaled by a time-dependent coefficient in this conservation equation. This implies that the matter sector must involve some abelian symmetry; though at this stage we cannot identify its details as we would need more information about the matter sector (we will later see that locality requires it to be related to a space-time symmetry). 
On the other hand the vector component of the usual conservation of sources $\partial_\mu T^{\mu\nu}=0$ is rather complicated
\bea
&\partial_j& \tau_{ij} + {A \over H} \dot{p}_i = \left( I - \frac{AB}{H} \right)\! \frac{ \partial_i \ddot{\rho} }{C \nabla^2} + \frac{ E - J }{C} \partial_i \rho \nonumber \\ 
&+& \frac{D-C}{\ks} \partial_i \nabla^2 \phi + \Bigg[ {AB\over H}\left( {\dot{B}\over B} - {\dot{H}\over H} \right) -\dot{I}\Bigg] {\partial \over \partial t} {\partial_i \over \nabla^2} \! \left( {\rho \over C} \right) \,\,\,\,\,\,\,\,\,\,\,\,\label{taueq}
\eea

{\em Enforcing locality}:
The basic principle which we are imposing is that of locality: no instantaneous signaling at a distance. However the previous equation will in general spoil this because there are several terms that involve inverse Laplacians. This means that even if the sources themselves ($\rho$, etc.) are initially local, they will evolve to ensure sources ($p_i$,\,$\tau_{ij}$) are non-locally distributed in space. This would provide a means to instantaneously signal to a distant observer by having the distant observer situated near one of these effectively non-local sources and then disturbing the source. Hence to maintain locality we need to enforce that these terms vanish. This requires the following set of conditions be satisfied
\beq
AB = IH \label{acond1},~~~~
{\dot{B} \over B} = {\dot{I} \over I} + {\dot{H} \over H},~~~~ 
D=C 
\eeq 
where the first condition here arises from requiring the coefficient of $\partial_i\ddot{\rho}/\nabla^2$ vanish, the second condition here arises from requiring the coefficient of $\partial_i(\rho/C)/\nabla^2$ vanish, and the third condition here arises from requiring the coefficient of $\nabla^2\phi$ vanish (this final condition is needed because $\phi$ itself is highly non-local; see ahead to eq. (\ref{phiA})). With these 3 conditions enforced, the conservation equation becomes local and nearly canonical. To make it appear as canonical as possible it is useful to define a shifted tensor source as 
\beq
\tau_{ij}=\tilde{\tau}_{ij}+{E-J\over D}\delta_{ij}\rho
\label{tildetau}\eeq 
By then inserting into eq.~(\ref{taueq}) with the above trio of conditions, we then obtain
\beq
\partial_j \tilde{\tau}_{ij} + {A \over H} \dot{p}_i = 0
\label{tildetauCons}\eeq	
So, in this theory, locality has enforced that the sources obey almost standard conservation laws. (We note that $\tilde{\tau}_{ij}$ couples linearly to the graviton, just as $\tau_{ij}$ does.) 

Having ensured that the sources themselves are local, we can now move to another test of locality. Since the graviton is massless (or even if it were very light) it will mediate a long ranged force between matter sources. Hence it can very easily lead to non-local interactions. One necessary condition to avoid this is that the exchange interaction between a pair of matter sources, with no external gravitons, is local. (Later we will comment on additional ramifications that can emerge from considering external gravitons.) So it is useful to have the inhomogeneous particular solutions for the gravitational fields
\bea			
&&	{h\over\ks} = \frac{ - \rho}{D\nabla^2} \label{hA}\\
&&	{\psi_i\over\ks} = \frac{p_i}{H \nabla^2} + \frac{B}{H} \frac{\partial_i}{\nabla^4} {\partial \over \partial t} \! \left( {\rho \over D} \right) \label{psiA}\\
&&	{\phi\over\ks} = \frac{-\tau}{2D \nabla^2} + \frac{E + L - 3J}{2D^2} \frac{\rho}{ \nabla^2}  \nonumber\\ 
&&+  \frac{1}{2D} {\partial \over \partial t} \bigg\lbrace \left( 3I\! - \!K \right) \left[ {\partial \over \partial t} \left( \frac{\rho}{D \nabla^4} \right) \right] \bigg\rbrace \label{phiA}\\
&&	{h_{ij}\over\ks} = \frac{\tau_{ij}}{\tilde{\square}} + \frac{\left( \partial_i \partial_j - \delta_{ij} \nabla^2 \right) \tau}{2 \tilde{\square} \nabla^2} + {1 \over \tilde{\square}} \Bigg( {A \over \nabla^2} {\partial \over \partial t} \! \left( \frac{\partial_{(i}p_{j)}}{H} \right) \nonumber\\
	&&+ {\delta_{ij} \over 2D}(E\! - \!J\! + \!L) \rho + \left( { 3J\! - \!3E \! - \!L \over 2D} \right) {\partial_i \partial_j \rho \over \nabla^2} \nonumber \\
	&&+ {\delta_{ij} \over 2 \nabla^2} {\partial \over \partial t} \left[ (I \! - \! K) {\partial \over \partial t} \! \left( \rho \over D \right) \right] + {\partial_i \partial_ j \over \nabla^4} \bigg\lbrace 2A {\partial \over \partial t} \left[ {B \over H}  {\partial \over \partial t} \! \left( {\rho \over D} \right) \right] \nonumber \\
	&&+ {1 \over 2}  {\partial \over \partial t} \left[ (K\! - \!3I)  {\partial \over \partial t} \! \left( {\rho \over D} \right) \right] \bigg\rbrace \Bigg)
	\label{gaugeFields}
\eea
where we have defined a wave operator
\beq
	\tilde{\square} \equiv K \partial_t^2 + \dot{K} \partial_t  - L\nabla^2. \label{box} 
\eeq
This unusual wave operator merits some discussion. First, though it involves functions of time through $K=K(t)$ and $L=L(t)$ it is still a linear operator, and therefore has a Green's function, which lets us continue using its inverse to simply mean convolution with that Green's function. Or, more simply, whatever its functional form, this operator has a matrix representation, and the operations of matrix multiplication and inversion are well defined. 

The corresponding tree-level graviton-exchange action between a pair of sources is in fact 1/2 of the interaction term $\Lex=\Lint/2=-\kappa\,h_{\mu\nu}\T^{\mu\nu}/4$. Then to determine under what conditions this exchange action is local, we can now write the interaction Lagrangian eq.~(\ref{Lint}) only in terms of the sources by using the solutions (\ref{hA}--\ref{gaugeFields}) to eliminate the fields from the interaction Lagrangian. We can then integrate by parts (in the spatial integrals) in the action to replace all divergences using the conservation equations (\ref{cons0}) and (\ref{tildetauCons}). Integrating by parts in time now introduces a great deal of complexity to the problem because each term in the action is cubic in functions of time, and the operator $\square$ in eq.$~$(\ref{box}) also contains functions of time. The general non-local exchange action contains a great many unique terms. To study this in full generality was only feasible with computer algebra software, such as Mathematica. We will describe its general form and present a few simple terms as concrete examples.

Schematically, the exchange action now contains terms of the form 
\bea
	\Lex \supsetsim &a_1&\![t] \,  \So \, \mathcal{D}(a_2[t] \So ) + a_3[t] \,  \So \, \mathcal{D}(a_4[t] \dot{\So}) \nonumber \\
	&+& a_5[t] \,  \So \, \mathcal{D}(a_6[t] \ddot{\So}) \label{example}
\eea
where $a_n[t] = a_n[A(t),B(t),\mathellipsis,L(t)]$ are functions of the coefficients in the action and their time derivatives, $\mathcal{D} = \mathcal{D}[\tilde{\square}^{-1},\nabla^2]$ is some differential operator involving at least one power of the inverse wave operator and various (positive or negative) powers of Laplacians, and $\So$ stands for any of the sources $\rho,\,p_i,\,\tau_{ij}$. Note in the static limit ($\dot{A}=\dot{B}=\mathellipsis=\dot{L}=0$), the second term would be a total derivative in the action, while the other two terms would not necessarily vanish. To enforce locality, for every distinct term in $\Lex$ that contains any inverse Laplacians the functions $a_n[t]$ must be set to zero or a constant to eliminate the non-local terms, making them identically zero or a total derivative as appropriate. In most cases each set of terms provides multiple possible constraint options to eliminate the non-local parts, such as in the first term of the schematic example eq.$~$(\ref{example}) where if this term were non-local, either $a_1$ or $a_2$ could be set to zero to eliminate it. Nevertheless, it turns out there is a unique set of conditions that eliminates all non-local terms from the exchange action. 

We emphasize that any terms in $\Lex$ that have inverse Laplacians would give rise to instantaneous signaling of one source to a distant source; which we wish to avoid in this work. This would be in sharp contrast to all known physical theories (general relativity, electromagnetism, etc) which avoid instantaneous signaling; and indeed avoid inverse Laplacians in the final exchange action.

Most of the sources can only couple to themselves, as represented in eq.$~$(\ref{example}) where each term $\sim \So \So$. The only possible mixed terms are $\sim \rho\, \tau$, and turn out to be among the least numerous. As a simple example, grouping non-local terms $\sim \rho \,\tau$, the exchange action contains
\beq
	\Lex \supset {\tau \over 2 C} {\left( I-K \right) \ddot{\rho} \over \tilde{\square} \nabla^2} + {I \ddot{\rho} \over 2 C}{\tau \over \tilde{\square} \nabla^2} - {\rho \over 2 C} {K \ddot{\tau} \over \tilde{\square} \nabla^2} \label{rhotauA}
\eeq
from which we find locality requires $I=K$ in order to remove the first term as it is proportional to $1/\nabla^2$. We also find $\dot{K}=0$ is a necessary condition for locality, rendering $I=K=const$. Note, however, this does not completely eliminate every term in eq.$~$(\ref{rhotauA}), just the non-local ones; integrating by parts in time generates local terms higher order in $\square^{-1}$. The full set of necessary and sufficient conditions to eliminate all non-local terms from the tree-level exchange action in the theory of this subsection are found to be
\bea
&&	A=B=c_1 \, , ~~~ C=D=c_2 \, ,~~~ I=K=c_3 \, , ~~~ H={c_1^2 \over c_3}  \nonumber\\
&&	2E(t)=J(t)+L(t) \, , ~~~ \dot{E} = \dot{J} = \dot{L} = c_4  \label{Aconds}
\eea
where $c_1,\,c_2,\,c_3,\,c_4$ are constants (independent of time); $c_1,\,c_2,\,c_3$ are dimensionless, while $c_4$ has units of inverse time. We see that locality imposes that most of the coefficients obey the standard relations in the Poincar\'e invariant theory of general relativity. In particular, most of the coefficients are required to be constants. Moreover the conservation equations (\ref{cons0},\,\ref{tildetauCons}) are now simple because the coefficients in those equations are required to be time independent. In fact building $\T^{\mu\nu}$ out of the sources $\rho,~p_i$ and $\tilde{\tau}_{ij}$, we can easily rescale $\rho$ and $p_i$ by a constant to write $\partial_\mu\tilde{\T}^{\mu\nu}=0$, so the theory is required to have a canonical conserved source.

The new feature that departs from the Poincar\'e invariant theory are the functions $E,~J,~\text{and }L$ which can be linear functions of time, and must have the same time dependence. Writing $J = J_0 + c_4 t$ etc., enforcing the conditions (\ref{Aconds}) leaves the exchange action
\bea
&&\frac{8\Lex}{\kappa^2} = \tau_{ij} \frac{\tau_{ij}}{\square} - \! \frac{\tau}{2} \frac{\tau}{\square} + \! {\{\rho \over 4 c_2} \frac{a_1 \tau\}}{\square}  
+ \! \frac{\rho}{8 c_2^2} \frac{a_2 \rho}{\square} - \! {2 c_3 \over c_1^2} \,p_i \frac{L p_i}{\square} \nonumber \\
&&+ {c_3 c_4 \over c_2} \rho  \left[ {\dot{\tau} \over \square^2}  + c_4 {\nabla^2 \tau \over \square^3} \right] -{c_3 c_4 \over  c_2^2} \rho  \left[  c_4 {\rho \over \square^2} + {(L-3J) \dot{\rho} \over 2 \square^2}  \right] \nonumber \\
&&+{c_3 c_4^2 \over 2 c_2^2} \rho {(3J-5L) \nabla^2 \rho \over \square^3}                         -{2 c_3^2 c_4\over c_1^2} p_i {\dot{p}_i \over \square^2}                \label{localLintA}
\eea
where $a_1=3L-J$, $a_2= 18 J L - 3 J^2  - 11 L^2$, $a_3 = 3J-5L$, $\{\rho\,a_1\tau\}/\square\equiv\rho(a_1\tau/\square)+\tau(a_1\rho/\square)$, and the wave operator with $K$ constant has become $\square \equiv K \partial_t^2 - L\nabla^2$. This is now completely local and the first line has a similar form to the GR action eq.$~$(\ref{Lint_GR}), but with coefficients $a_1$ linear in time and $a_2$ quadratic in time. We are left with six unspecified parameters, $c_1,c_2,c_3,c_4,J_0$ and $L_0$. In the static case ($c_4=0$) the two symmetric $\rho\, \tau$ terms on the first line combine and the higher order $\square^{-1}$ terms vanish, in agreement with our previous work on the static case. However, with $c_4 \neq 0$, we are left with a more general -- explicitly time dependent -- theory.

Even in the static limit, it is not immediately obvious how eq.$~$(\ref{localLintA}) reduces to eq.$~$(\ref{Lint_GR}). As we remarked in \cite{part2}, if we rewrite the exchange action in terms of the conserved sources and rescale $\rho \to (c_2/L) \rho$, in the static case we then recover eq.$~$(\ref{Lint_GR}) exactly. However, this does not generalize to the present case with $L(t)$ a function of time. Expressing the exchange action (\ref{localLintA}) in terms of the conserved source $\tilde\tau_{ij}$ (from eqs.$~$(\ref{tildetau},\ref{tildetauCons})) we obtain
\bea
\frac{8\Lex}{\kappa^2} &=& \tilde{\tau}_{ij} \frac{\tilde{\tau}_{ij}}{\square} - \frac{\tilde{\tau}}{2} \frac{\tilde{\tau}}{\square} + {\{\rho \over 2 c_2} \frac{L \tilde{\tau}\}}{\square} 
+ {\rho \over 2 c_2^2} {L^2 \rho \over \square} - {2 c_3 \over c_1^2} p_i {L p_i \over \square} \nonumber \\
&+& {c_3 c_4 \over  c_2} \left(  \rho {\dot{\tau} \over \square^2} + c_4 \rho {\nabla^2 \tau \over \square^3}  \right) + {c_3 c_4 \over c_2^2} \rho { L \dot{\rho} \over \square^2} + {c_3^2 c_4^2 \over  c_2^2} \rho {\ddot{\rho} \over \square^3} \nonumber \\
&-& 2{c_3^2 c_4 \over c_1^2} p_i {\dot{p}_i \over \square^2}            \label{localLint2}
\eea
In this form $J(t)$ drops out, so there are really only 5 constants here: $c_1,c_2,c_3,c_4,$ and $L_0$ (we may, as in the next section, label $L_0 \equiv c_5$). 
Unlike in the static case, factors of $t$ (from $L=L_0+c_4 t$) can only appear next to source terms inside the $\square^{-1}$. To appear to the left in any of these terms, outside the $\square^{-1}$, requires some integration by parts in time, leaving only $\sim \dot{L} = c_4$ outside the $\square^{-1}$. Thus in the two terms symmetric in $\rho$ and  $\tilde{\tau}$, in one term $L$ comes with the $\rho$ and in the other $L$ comes with the $\tilde{\tau}$. This means with $L(t)$ a function of time we cannot rescale $\rho$ to make the first line look more like eq.~(\ref{Lint_GR}) as we could in the static case. Our interpretation of the above result will come in Section \ref{Discussion}.
	
\subsection*{Theory B: Constraint from Equations of Motion}\label{grcase2}

In the above section we cut down to 2 degrees of freedom of the graviton by imposing constraints directly on the fields in eq.~(\ref{gauge}). A secondary approach is to impose some conditions on the pre-factor functions so that the equations of motion include constraints that cut down the degrees of freedom directly. By studying the full theory with all 12 unknown functions $A(t),\mathellipsis,L(t)$, and using the general equations for the non-conserved sources eq.~(\ref{sigmaEq}) and (\ref{wEq}), we can impose the constraints on functions
\bea
	&& A=B=c_1 \, , ~~~~~ C=D=c_2 \, , ~~~~~ I=K={A^2 \over H} ~~~~~ \label{Bconds1}
\eea
It is simple to show that this now fixes $\partial_i \psi_i$ and $\partial_i h_{ij}$ by the equations of motion, leaving the theory with the desired
2 degrees of freedom. In particular the classical equations of motion enforce that they are given by the sources as
\bea
\partial_i \psi_i &=& \frac{\ks \, \sigma}{(H-F)\nabla^2} \label{divpsi}\\
\partial_i h_{ij} &=& {\ks \over (G-L) \nabla^2} \bigg[ \left( {2E-G-J \over 2G+J-2E-L} \right) {\partial_j q \over \nabla^2} + w_j \nonumber \\
 &-& {K \over A} {\partial_j \dot{\sigma} \over \nabla^2} + \left( {\dot{F} \over F} - {\dot{H} \over H} \right) {AF \over H(H-F)} {\partial_j \sigma \over \nabla^2} \nonumber \\
 &+& {J-E \over D} \partial_j \rho + {\dot{K} \over A} p_j \bigg] \label{divhij}
\eea
In the expression for $\partial_i h_{ij}$ we have introduced the following function of the sources
\bea
q%[\T^{\mu\nu}] 
&\equiv& \partial_j w_j - \frac{A}{H} \dot{\sigma} + \frac{J-E}{ D} \nabla^2 \rho - {\dot{K} \over D} \dot{\rho} \nonumber \\
&+&\left[ \left( {\dot{F} \over F} - {\dot{H} \over H} \right) {AF \over H(H-F)} + {\dot{K} \over A} \right]\! \sigma   \label{qdef}
\eea
which gives a nice form for the second derivative
\beq
\partial_i \partial_j h_{ij} = \left( {\ks \over 2G-2E+J-L} \right) \!{q \over \nabla^2}
\eeq

{\em Enforcing locality}:
As in the previous section, we can now solve the equations of motion to obtain the inhomogeneous solutions for the fields, given in Appendix \ref{A}. We again use these to write the exchange action only in terms of the sources, integrating by parts to replace divergences using the definitions of $q$, $\sigma$, and $w_i$. Once again, because of the cubic dependence on functions of time, integrating by parts to simplify under the fewest independent variables expands the total number of terms enormously. The general form is similar to that described in eq.~(\ref{example})

Now that we have general $\sigma$ and $w_i$ (as opposed to the restricted form in Theory (A)), there are more possible mixed terms than only $\sim \rho \, \tau$; there are also terms with $ \tau \, \sigma$, $\rho \, \sigma$, $\rho \, \partial_i w_i$, $\tau \, \partial_i w_i$, and $p_i w_i$. It is advantageous to first look at the $w_i w_i$ and $\sigma\,\sigma$ terms involving inverse Laplacians to see if locality requires some constraint that simplifies all these other terms involving $\sigma$ and $w_i$. Looking for nonlocal terms in $w_i w_i$ and $\sigma\,\sigma$, the exchange action contains
\bea
	\Lex \supset 2 \sigma {1 \over \tilde{\square} \nabla^2} \left[ {FKL \, \sigma \over H\left( c_1^2-FK \right)} \right] + 2 w_i {1 \over \tilde{\square} \nabla^2} \left[ {G \, w_i \over G-L} \right] \,\,\,\,\,\,\,\,\,\,
\eea
plus terms higher order in $\nabla^{-2}$ and derivatives of $\sigma$ and $w_i$. If $\sigma$ and $w_i$ are general functions, then the theory is immediately non-local. To eliminate these terms, the only possible constraints on the coefficients $A(t),\mathellipsis,L(t)$ would impose some of these vanish (e.g., $G(t)=0$), leading to an overly trivial theory, which we decline to do. The most direct way to avoid this problem is to impose that $\sigma$ and $w_i$ vanish, i.e., $\sigma=w_i=0$. In previous work \cite{part2} we allowed for $\sigma$ and $w_i$ to themselves be given by derivatives of other local functions in order to maintain locality by having the Laplacians cancel out. But this only leads to ultra-local terms in the action, which were found to be associated with a completely decoupled, and hence irrelevant, sector. We will ignore this here as it does not represent a real modification to GR. Instead by focussing on the important case of only coupled sectors, we are forced by locality to impose $\sigma=w_i=0$ (we can also have $w_i=(E-J)\delta_{ij}\rho/D$, but we can easily shift from $\tau_{ij}$ to $\tilde{\tau}_{ij}$ to eliminate this, as we did earlier in eq.(\ref{tildetau}); so we can ignore that without loss of generality). 

Thus locality imposes that the sources are conserved and the theory is now similar in structure to Theory (A) of the previous section, though the coefficients $F$ and $G$ are still present. 
The full set of conditions to make the exchange action of Theory (B) local are
\bea
	&&\sigma = 0 \, , ~~~~ w_i = 0 \, , ~~~~ I=K=c_3 \, , \nonumber \\
	&& E=G=J=L=c_5+c_4 t, ~~~~ F=H=c_1^2/c_3 \label{Bconds2}
\eea 
Again we see that the static case is recovered if $c_4=0$ in agreement with Ref.~\cite{part2}. For the time dependent case with $c_4 \neq 0$, we have the conditions of eq.$~$(\ref{Bconds2}). So we are left with 5 parameters of the theory, as in Theory (A). 
We therefore recover the same local action eq.\,(\ref{localLint2}) with the same number of parameters.
(We note that since $F=H$ and $G=L$ here, it is compatible with eqs.~(\ref{divpsi},\ref{divhij}), since we also have $\sigma=w_i=0$. This is clearest seen by multiplying both sides of eq.~(\ref{divpsi}) by $(H-F)$ and eq.~(\ref{divhij}) by $(G-L)$).

%The pre-factor function $F$ does not explicitly appear in any of the conditions for locality, so it naively appears $F$ can be any function of time. Let us now consider this more carefully. With eq.$~$(\ref{Bconds2}) the constraint equations for $\partial_i \psi_i$ and $\partial_i h_{ij}$, (\ref{divpsi}) and (\ref{divhij}), become
%\bea
%\partial_i \psi_i = 0\,,~~~~~ \partial_i h_{ij} = 0 \label{zerodiv}
%\eea
%In particular, the requirement $\sigma=0$ forces $\partial_i \psi_i=0$. As seen in eq.\,(\ref{divpsi}), this is equivalent to setting $F=H=c_1^2/c_3$, so in fact $F$ is a constant and can be written in terms of the already enumerated parameters. In Theory (A) we demanded eq.\,(\ref{zerodiv}) as our starting point, eliminated $F$ and $G$ from the theory. Theory (B) allows a more general setup, but locality requires us to enforce these constraints anyway. 

\section{Discussion}\label{Discussion}

After enforcing locality, we see that both theories became identical, so we recover the same exchange action. That is, imposing all conditions in eq.~(\ref{Bconds2}), the now local tree-level exchange action of Theory (B) is exactly as in Theory (A) after writing the exchange action in terms of the conserved sources, eq.$~$(\ref{localLint2}). We believe this is true for any particular construction that follows from the above class of theories outlined in Sec.\,\ref{spin2theories} (we can also have decoupled sectors, as we elaborated on in \cite{part2}, but this is of little interest here). 

If we set the coefficient $c_4=0$ we recover general relativity exactly (at this order). But with $c_4 \neq0$ we have a theory of spin 2 with two degrees of freedom that explicitly breaks time translation invariance, while maintaining locality in tree-level graviton exchange between conserved matter sources, and is therefore a deviation from general relativity.

\subsection*{Gauss-Bonnet}

Analyzing only the exchange action as we have in the previous two sections lets us identify non-local interactions, but obfuscates the structure of the theory. Returning to the full Lagrangian, after enforcing locality by eq.~(\ref{Bconds2}) what we are left with is the Lagrangian of eq.~(\ref{General}) with all coefficients constant, plus a term $\propto c_4$ that breaks time translation symmetry. If we set $c_4 =0$, we recover time-translations, and we can also absorb the constant coefficients by a redefinition of the fields, allowing us to recover GR with its full Poincar\'e symmetry. In the case of interest, with $c_4 \neq 0$, the theory breaks time translation symmetry and in turn the Lorentz boost symmetry of the full theory. 

For ease of notation and without loss of generality, we set all $c_1=c_2=c_3=c_5=1$. This leaves only $c_4$ as the adjustable parameter of the theory. We can then write the full Lagrangian of the theory (\ref{General}) as
\bea
\Lag &=& \LGR^{(2)} + c_4 t  \bigg( {1 \over 2} \partial_i h \partial_i h  - {1 \over 2}\partial_k h_{ij} \partial_k h_{ij} \nonumber \\
&&~~~~~~~~~~~~+ \partial_j h_{ij} \partial_k h_{ik} -  \partial_i h \partial_j h_{ij} \bigg) \label{finalLint}	
\eea
Let us compare this to the Gauss-Bonnet action \cite{Lovelock:1971yv} with an inserted pre-factor function $f(t)$
\bea
\Delta\LGB = f(t) \sqrt{-g} \left( R^2 - 4R_{\mu\nu}R^{\mu\nu} + R_{\mu\nu\rho\sigma}R^{\mu\nu\rho\sigma} \right) \label{GB} ~~~~
\eea 
At the leading order in $h_{\mu\nu}$, to which we are working, and after performing some integrations by parts, we can identify 
\beq
\ddot{f} = - c_4 \, t \label{fcond}
\eeq
That is, the additional terms in eq.~(\ref{finalLint}) are nothing other than the Gauss-Bonnet action with the function $f(t)$ determined by eq.$~$(\ref{fcond}).

If $f$ is a constant, the action eq.$~$(\ref{GB}) is a surface term in four dimensions, so it is usually not included in GR even though this particular sum of curvature squared terms satisfies the same symmetries as the Einstein-Hilbert action. While we did not begin our analysis by explicitly adding the Gauss-Bonnet action with a function of time $f(t)$, these terms are contained in the general action of eq.~(\ref{General}) to quadratic order in $h_{\mu\nu}$. After enforcing locality we recover Einstein-Gauss-Bonnet gravity with $f(t)$ related to the graviton speed $v_g^{\,2}=L(t)$ by eq.~(\ref{fcond}). 

Since $\Delta \LGB$ is itself invariant under the linearized gauge transformations $h_{\mu\nu} \to h_{\mu\nu} + \partial_{(\mu}\alpha_{\nu)}$ and both Theories (A) and (B) end up with a conserved source, $\partial_\mu \tilde{\T}^{\mu\nu}=0$, the full action of eq.$~$(\ref{finalLint}) enjoys the linearized gauge redundancy of GR at this leading order. One can then gauge fix in any desired fashion (e.g., as in Theory (A)) and thus obtain eq.$~$(\ref{localLint2}) for the tree-level exchange action between matter sources. 
%So while we did not assume the starting action (\ref{General}) carried any gauge redundancy, we see that it is demanded by locality (at this order). 

So the general conceptual features of our final results are: the imposition of locality (avoiding instantaneous signaling) is so strong it forces our theory into a familiar gauge redundant form, even though we did not assume gauge redundancy as a starting point in eq.~(\ref{General}). In particular, in addition to the quadratic Einstein-Hilbert term, which is well known to be gauge invariant, we have also found an allowed Gauss-Bonnet term with a pre-factor that depends on time. This in fact is the most general action at the order we are working (albeit with a particular function of time) which is gauge invariant. 
A general conceptual feature of the sources $\T^{\mu\nu}$ will be discussed in the final sub-section.

\subsection*{ Broken Symmetries at this Order}

The general local action we find for spin 2 with two degrees of freedom, eq.$~$(\ref{finalLint}), has the linearized gauge invariance of GR, but the presence of the explicit function of time $f(t)$ prevents time translation symmetry, and its relation to the graviton speed by eq.$~$(\ref{fcond}) (i.e., $v_g^{\,2}=L(t)$, $\ddot{f} = -\dot{L}$) prevents us from recovering the usual Lorentz boost invariance. We do not have a simple intuition as to why locality of spin 2 prefers to include the Gauss-Bonnet terms in the action with a prefactor $f(t)$ that can be at most cubic in time, when any function of time would maintain gauge invariance at this order. 

In previous work \cite{Hertzberg:2017abn} some of us found for any non-trivial $f(t)$ the theory of eq.$~$(\ref{finalLint}) allows for superluminal gravitons. This is still true here, but in this work we demanded only a general notion of locality, that there is no instantaneous signaling. So the theory is local in this sense, though allows for gravitons with a time-dependent speed which can in general be superluminal, i.e., it can be faster than the particle speeds of the matter sector, including photons.

Non-zero $c_4$ characterizes deviations from Poincar\'e invariance. To be compatible with observations the value of $c_4$ (which has units of inverse time) must be very small. As an example of a bound, the recent LIGO observations of merging neutron stars showed that the graviton's speed matched the photon's speed to an accuracy of $|v_g-1|\lesssim 10^{-15}$ \cite{Monitor:2017mdv} (we set the speed of light to 1 here). Since the photons and gravitons travelled from this event for $t_*\sim10^8$\,years to earth, we can translate this into the direct bound on the coefficient $c_4$ of
\beq
|c_4|\lesssim |v_g-1|/t_* \sim 10^{-23}\,\mbox{years}^{-1}
\eeq
Since $c_4$ is dimensionful it is not obvious what its {\em a priori} expected value is. On the other hand, since this is such a small number in most fundamental units, it suggests that its true value may very well be exactly zero.

\subsection*{Derived Symmetries and Higher Order}

One of our primary findings is that locality requires the conservation of matter sources $\partial_\mu\tilde{\T}^{\mu\nu}=0$. Since there is only one non-trivially conserved 2-index tensor, then this must be the energy-momentum tensor $\tilde{\T}^{\mu\nu}=T^{\mu\nu}$. As we detailed at the end of \cite{part2}, the existence of a conserved symmetric energy-momentum tensor relies both on Lorentz and time translation invariance. Hence we have found that locality enforces those symmetries are obeyed by the matter sector.

On the other hand, our final action allows for the breaking of time translations in the gravity sector. At the above level of analysis, this is not an inconsistency because we have only worked to leading order; in particular, we only need the {\em matter sector} to exhibit time translations for $\partial_\mu T^{\mu\nu}=0$, and this is true so long as we ignore backreaction from the graviton. Nevertheless this does suggest that if one were to demand locality at higher order, then it is possible that the time translation violation, as measured by the parameter $c_4$, would in fact have to vanish. 

Moreover, one could reasonably anticipate this as follows: so far we have only allowed the external sources to be part of the matter sector and we found that such a matter sector must exhibit time translations in order to maintain locality. In addition, however, one should also consider the external sources to be gravitons. In this case, it seems plausible that a similar analysis would demand that the energy-momentum tensor of the graviton itself needs to be conserved, which would then demand time translations in the graviton sector, giving $c_4=0$ and the full Poincar\'e symmetry of general relativity. Another hint is the following: we have remarked that locality has enforced that the final action of eq.~(\ref{finalLint}) is gauge invariant to this leading order. It is plausible that at higher order, locality will again demand gauge invariance; this would presumably require $f$ to be a constant to recover the usual non-linear diffeomorphism invariance of general relativity. But we do not have a proof of these issues. 

An important direction for future work is to include mass terms and single derivative terms in the starting action. Although we could show that mass terms needed to vanish in order to maintain locality in the time translation invariant  case of \cite{part2} (and the single derivative terms could then be converted by field re-definitions into higher dimension operators), they may play an important role in the time dependent case. This is especially important in that such terms are known to exist when expanding around an FRW background in ordinary general relativity. We leave this important issue for future work.

%\vspace{0.5cm}
 
\section*{Acknowledgments}
We would like to thank McCullen Sandora for very helpful input. 
MPH is supported in part by National Science Foundation grants No. PHY-1720332 and No. No. PHY-2013953.

%\newpage
\appendix 

\begin{widetext}
	\section{Equations of motion in Theory B (Sec. \ref{grcase2})} \label{A}
	In this appendix we report on the inhomogeneous particular solutions to the equations of motion (\ref{equations of motion1}--\ref{equations of motion}) with the conditions of Theory (B) to cut down to two degrees of freedom, eqs.$~$(\ref{divpsi}) and (\ref{divhij}).
	
 	\bea
 		\phi &=& {- \tau \over 2C \nabla^2} + \dt \left[ \left( {3I-K \over 2C \nabla^4} \right) \dt \left( {\rho \over C} - {q \over (2G-2E+J-L) \nabla^2 } \right) \right] + \left( {E-3J+L \over 2C^2 } \right) {\rho \over \nabla^2} \nonumber \\
 		&+& \left( { 2G- 4E+3J-L \over 2G-2E+J-L } \right) { q \over 2C \nabla^4} + {2A \over C} \dt \left[ {\sigma \over (H-F) \nabla^4} \right]     \\
 		\psi_i &=& {p_i \over H \nabla^2} + {F \over H(H-F) } {\partial_i \sigma \over \nabla^4} + {A \over H} \dt \Bigg\lbrace  {1 \over (G-L) \nabla^4} \left[ w_i - {K \over A} {\partial_i \dot{\sigma} \over \nabla^2} + \left({\dot{F} \over F} - {\dot{H} \over H} \right) {A F \over H(H-F)} {\partial_i \sigma \over \nabla^2} + {\dot{K} \over A} p_i \right]  \nonumber \\
 		&+& {A \over C} \left( 1+ {J-E \over G-L} \right) {\partial_i \rho \over \nabla^4} - {A \over G-L} {\partial_i q \over \nabla^6}           \Bigg\rbrace     \\
 		h_{ij} &=& {1\over \tilde{\square}} \Bigg\lbrace \tau_{ij} + \left( {\partial_i \partial_j \over \nabla^2} - \delta_{ij} \right) {\tau \over 2} + {\delta_{ij} \over 2} \left({E+L-J \over C} \right) \rho + {\partial_i \partial_j \over 2 \nabla^2} \left[ { L(L-3J)-G(J+L)+E(G+3L) \over C(G-L) } \right] \rho \nonumber \\
 		&+&{\partial_i \partial_j \over \nabla^4} \dt \left[ 2K \dt \left( \left\lbrace 1-{J-E \over G-L} \right\rbrace {\rho \over C} \right) -  {K \dot{\rho} \over C} \right] + {\delta_{ij} \over 2} \left( {2G-2E+3J-L \over 2G-2E+J-L} \right) {q \over \nabla^2} \nonumber \\
 		&+& {\partial_i \partial_j \over 2 \nabla^4} \left[{ G(2G-2E+J) + 3L(G-2E+J) - L^2 \over (G-L)(2G-2E+J-L)} \right] q + {\partial_i \partial_j \over \nabla^6} {\partial^2 \over \partial t^2} \left[ \frac{K (4E-3G-2J+L) \, q}{(G-L) (2G -2E+J-L)} \right] \nonumber \\
 		&+& {2 \partial_i \partial_j \over \nabla^4} \left[ A \dt \left( {F \, \sigma \over H(H-F)} \right) + {GK \, \dot{\sigma} \over A(G-L) } - \left( {\dot{F} \over F} - {\dot{H} \over H} \right) { AFG \, \sigma \over H(H-F)(G-L) } - A \dt \left( {\sigma \over H-F} \right) \right] \nonumber \\
 		&+& {2 \partial_i \partial_j \over \nabla^6} \dt \left[ K \dt \left( {1 \over G-L} \left\lbrace {-K \dot{\sigma} \over A} + \left({\dot{F} \over F} - {\dot{H} \over H} \right) {AF \, \sigma \over H(H-F)} \right\rbrace \right) \right]  + A {\partial_{(i} \over \nabla^2} \dt \left( {p_{j)} \over H} \right) \nonumber \\
 		&+& {A^2 \partial_{(i} \over \nabla^4} {\partial^2 \over \partial t^2} \left[ {1 \over G-L} \left( w_{j)} + {\dot{K} \over A} p_{j)} \right) \right] - {G \partial_{(i} \over (G-L) \nabla^2} \left( w_{j)} + {\dot{K} \over A} p_{j)} \right) \Bigg\rbrace 
 	\eea
 	where as above $\tilde{\square} \equiv \partial_t (K \partial_t) - L \nabla^2$, and we can take the trace to obtain
 	\bea
 		h &=& {-\rho \over C \nabla^2} + {q \over (2G-2E+J-L)\nabla^4} 
 	\eea
		
\end{widetext}

\end{document}